\documentclass[prl,amsmath,superscriptaddress,citeautoscript,twocolumn,
showpacs,floatfix]{revtex4}
\usepackage{graphicx}

\usepackage{amssymb}
\usepackage{color}
\usepackage{epsfig}

\begin{document}

\def\sigmav{{\mbox{\boldmath{$\sigma$}}}}

\title{Partial decoherence in a qubit coupled to a control register subject  to telegraph noise}
\author{Amnon Aharony}
\email{aaharony@bgu.ac.il}
\altaffiliation{Also at Tel Aviv University, Tel Aviv 69978,
Israel}

\affiliation{Department of Physics and the Ilse Katz Center for
Meso- and Nano-Scale Science and Technology, Ben-Gurion University, Beer
Sheva 84105, Israel}

\author{Ora Entin-Wohlman}
\altaffiliation{Also at Tel Aviv University, Tel Aviv 69978,
Israel}

\affiliation{Department of Physics and the Ilse Katz Center for
Meso- and Nano-Scale Science and Technology, Ben-Gurion
University, Beer Sheva 84105, Israel}

\author{Sushanta Dattagupta}
\affiliation{ Indian Institute of Science Education and
Research-Kolkata, Mohanpur 741252, India}

\date{\today}
\begin{abstract}

A qubit (containing two quantum states, 1 and 2), is coupled to a
control register (state 3), which is subject to telegraph noise.
We study the time evolution of the density matrix $\rho$ of an
electron which starts in some coherent state on the qubit. At
infinite time, $\rho$ usually approaches the fully decoherent
state, with $\rho^{}_{nm}=\delta^{}_{nm}/3$. However, when the
Hamiltonian is symmetric under $1\leftrightarrow 2$, the element
$\rho^{}_{12}$ approaches a non-zero real value, implying a
partial coherence of the asymptotic state. The asymptotic density
matrix depends only on ${\rm Re}[\rho^{}_{12}(t=0)]$. In several
cases, the information stored on the qubit is protected from the
noise.

\end{abstract}
\pacs{03.67.-a, 05.40.-a, 03.65.Yz, 03.67.Lx}

\maketitle

Quantum computation operates on information  stored in ``qubits",
which are  superpositions of two basic quantum states 
\cite{bennett}.   Clearly, quantum computation requires the
stability of the quantum state stored on each qubit, and therefore
it can be used only while this state remains coherent
\cite{zurek}. Interactions between  qubits and their environment,
including input-output measurement devices, can cause decoherence
which destroys the information stored in the qubits. Attempts to
avoid decoherence have led to studies of decoherence-free
subspaces, within which the quantum state remains protected
\cite{Lidar}. The states in such subspaces are practically
decoupled from the sources of the decoherence, due to symmetries
of the system.

Several proposals \cite{beige,kulik} to reduce the decoherence of
the single qubit states use {\it three} basic states: two
degenerate states ($|1\rangle$ and $|2\rangle$) forming  the
qubit, and the third state ($|3\rangle$) representing a control
register. Models include an electron in a three-state atom
\cite{beige}, or in three  single-level quantum dots \cite{divin}.
Below we refer to these states as sites in a tight binding model
on a ring. The general Hamiltonian for such a 
system contains ``site" energies $\epsilon^{}_n$ and ``hopping"
matrix elements $J^{}_{nm}$,
\begin{align}\label{H}
 \mathcal{H}^{}_0=\sum_n\bigl (\epsilon^{}_n|n\rangle\langle n|-(J^{}_{nn+1}|n\rangle\langle n+1|+{\rm h.c.})\bigr )~,
\end{align}
where we identify $n=4$ with $n=1$. In the non-stochastic case,
one observes Rabi oscillations of the electron wave function, with
frequencies equal to differences between the three eigen-energies
of ${\cal H}^{}_0$. These oscillations are expected to decay when
the system is coupled
 to the environment, due to decoherence.

We represent the latter coupling  by a stochastic telegraph noise
\cite{datta}, which affects only the control register:
$\epsilon^{}_3$ follows a Markov process or a two level jump
process, and is given by $\epsilon^{}_3=\epsilon-f(t)\zeta$, where
$f(t)$ jumps randomly from $1$
 to $-1$ (or from $-1$ to $1$) with rate
$w^{}_{-+}$ (or $w^{}_{+-}$). These jumps in $f(t)$ result from a
contact with some noise source, so that the probabilities
$p^{}_\pm$ to find $f=\pm 1$ obey a Boltzmann distribution
with energy gap $\Delta E$ at temperature $T$, $
p^{}_+=1-p^{}_-=1/[1+e^{-\Delta E/k^{}_BT}]$. Detailed balance
implies $p^{}_{-}w^{}_{+-}=p^{}_{+}w^{}_{-+}$, and therefore the
jump rates can be written as \begin{align} w^{}_{\pm \mp}=\lambda
p^{}_{\pm}\ , \label{wp} \end{align} where
 $ \lambda =w^{}_{+-} + w^{}_{-+}$.
Such a fluctuating site energy can be envisaged to occur because
of noisy gate voltages or because of the coupling to a device
(e.g. a point contact) which measures the occupation on that site
\cite{gur1}. Being based on symmetry considerations, our results
may be expected to hold in other models as well.

 Writing the quantum state of the system as
$|\psi\rangle=\sum_n c^{}_n|n\rangle$, the density matrix
$\rho=|\psi\rangle\langle\psi |$ has elements
$\rho^{}_{nm}=c^{}_nc_m^\ast$. Decoherence  is usually associated
with {\it dephasing}:  the noise generates random fluctuations in
the relative phases of $c^{}_n$ and $c^{}_m$, causing the decay of
the average off-diagonal matrix elements. Full decoherence
 corresponds to a  fully ``mixed" state, where all the off-diagonal
elements of $\rho$ vanish  and all the three basis states (in any
representation) have equal occupation probabilities (in our case
$\rho^{}_{nn}=1/3$).  In contrast, {\it relaxation} describes the
decay of the system towards a ``pure" state, which is a coherent
eigenstate (the ground state  at zero temperature) of the
non-stochastic Hamiltonian \cite{gur2}.   Possible measures of
decoherence include
\begin{align}
{\cal D}\equiv 1-{\rm Tr}(\rho^2)\ ,\label{DD}
\end{align}
(one always has ${\rm Tr}\rho=1$) or the von Neumann entropy,
\begin{align}
{\cal S}=-{\rm Tr}(\rho\log\rho)\ .\label{SS}
\end{align}
In these expressions, $\rho$ represents a certain average over the
 stochastic noise (see below).
Both measures vanish for a ``pure" state, where the eigenvalues of
$\rho$ are equal to $1$ and two zeroes,
 and have maxima ($2/3$
and $\log 3$, respectively) for the fully decoherent state.
Intermediate values of these measures require that off-diagonal
matrix elements of $\rho$ differ from zero, representing {\it
partial decoherence}.   Such deviations from full decoherence, as
found below, can probably be interpreted as
 a mixture of decoherence and relaxation \cite{gur2}.

Below we find the stochastic evolution in time of $\rho$,
beginning with an initial coherent state in which the electron is
on the qubit [i.e. $R^+=1$, where $R^\pm\equiv
\rho^{}_{11}(t=0)\pm \rho^{}_{22}(t=0)$]. For arbitrary parameters
of the non-stochastic Hamiltonian ($\epsilon^{}_n,~J^{}_{nm}$) and
of the noise ($\zeta,~w^{}_{\mp\pm}$), our system develops full
decoherence. However, when the Hamiltonian is symmetric under the
interchange of the two qubit states, $1 \leftrightarrow 2$, then
the system maintains {\it partial} decoherence:  although
$\rho^{}_{13}$ and $\rho^{}_{23}$ (i.e., the  matrix elements
which mix the qubit and the register states) approach zero, the
matrix element $\rho^{}_{12}$  can approach a non-zero real value,
implying an entanglement of the final qubit states $|1\rangle$ and
$|2\rangle$. Furthermore, these states attain asymptotically equal
populations $\rho^{}_{11}=\rho^{}_{22}=(1-\rho^{}_{33})/2$.
Surprisingly, these asymptotic values of $\rho$ depend {\it only}
on the initial value of the parameter $y={\rm
Re}[\rho^{}_{12}(t=0)]$, which contains information on the initial
entanglement of the two qubit states, and {\it not} on any of the
other parameters (of the Hamiltonian, the noise or the initial
state)! A fully coherent initial state would have
$|\rho^{}_{12}|^2=\rho^{}_{11}\rho^{}_{22}$, and therefore $y^2\le
[1-(R^-)^2]/4$. In particular, the limiting values $y=\pm 1/2$
(and $R^-=0$) correspond to the ``bonding" and ``anti-bonding"
states $(|1\rangle\pm|2\rangle)/\sqrt{2}$, which are shown below
to be protected from the noise, forming  an example for a
decoherence-free subspace.

Here we discuss the simplest symmetric case, which contains only
hopping  between $|1\rangle$ or $|2\rangle$ to $|3\rangle$:  we
replace ${\cal H}_0$ by ${\cal H}={\cal H}^{}_0-f(t){\cal V}$,
with
 ${\cal V}=\zeta |3\rangle\langle 3|$, and set
 $\epsilon^{}_1=\epsilon^{}_2=0,~\epsilon^{}_3=\epsilon,~J^{}_{12}=0,~J^{}_{13}=J^{}_{23}=J$.
 The $3\times 3$ density matrix
${\rho}(t)$ obeys the von-Neumann-Liouville equation,
$i\partial_t\rho(t)=[{\cal H},\rho(t)]$. It is convenient to
rewrite this equation using the Liouville super-operators,
$i\partial_t\rho(t)={\cal H}^\times \rho(t)=\bigl [{\cal
H}^\times_{0}-f(t){\cal V}^\times(t)\bigr ] \rho(t)$, with the
obvious   definition ${\cal A}^\times \rho \equiv [\mathcal{A},
\rho(t)]$.

A treatment of the equation of motion with the underlying
stochasticity in $f(t)$ can be found in the literature on the
stochastic theory of lineshapes \cite{10}. Here we follow  Blume
\cite{11}, and replace each dynamic operator ${\cal O}(t)$ by a $2
\times 2$ matrix ${\widetilde{\cal O}}(t)$, such that $(
b|{\widetilde{\cal O}}(t)|a)$ represents the {\it conditional
average} of ${\cal O}(t)$, given that the stochastic variable
$f(t)$ had the value $a(=\pm 1)$ at time $t=0$ and the value $b$
(=$\pm 1)$  at time $t$. At the end one may average over the
stochastic process.  The choice of the averaging procedure depends
on the particular experiment. If one performs only one experiment,
with a single initial state of the noise, then the initial value
of $f(t)$ is fixed (say at $a$). The final value of $f(t)$ can
then be either $b=+1$ or $b=-1$, and thus an appropriate average
is
\begin{align}
 [{\cal O}]_{av,a}=\sum_{b=\pm 1} (
 b|\widetilde{\cal O}|a)~.\label{ava}
\end{align}
In many quantum measurements it is more appropriate to repeat the
measurement many times, on systems which are prepared in the same
way. In this case, it is necessary also to average over the
initial value of $f(t)$ (which equals $a=\pm 1$ with probability
$p^{}_a$), namely
\begin{align}\label{av}
 [{\cal O}]_{av}=\sum_{a,b=\pm 1}p^{}_a ( b|\widetilde{\cal O}|a) \equiv \sum_a p^{}_a [{\cal O}]_{av,a}~.
\end{align}
For finite  temperature $T$, i.e. $|\Delta p|<1$, where $\Delta
p=p^{}_+-p^{}_-$, our calculations yield similar qualitative
results for all these averages. In particular, all the averages coincide in the asymptotic limit $t\rightarrow\infty$.

We next construct a Markov chain of events in the time interval
$0$ to $t$, where the 
back and forth jumps of $f(t)$ are random and Poisson distributed
\cite{12}. The time evolution of $( b|\widetilde{\rho}(t)|a)$ is
now divided into two parts:
\begin{align}
\partial_t( b|\widetilde{\rho}|a)&=-i\bigl [\mathcal{H}^\times_{0}- b{\cal V}^\times\bigr ]( b|\widetilde{\rho}|a)\nonumber\\
&+w^{}_{b,-b}( -b|\widetilde{\rho}|a)-w^{}_{-b,b}(
b|\widetilde{\rho}|a)~.\label{dtrho}
\end{align}
The first term on the RHS applies if $f(t)$ remains unchanged at
time $t$ (i.e. equal to $b$). In this case, the time evolution
proceeds with the Liouville operator which corresponds to the
original Hamiltonian, with $\epsilon^{}_3=\epsilon-b\zeta$. The
last two terms arise if $f(t)$ flips exactly at time $t$, either
from $-b$ to $b$ (first term in the second line) or from $b$ to
$-b$ (last term). Equation (\ref{dtrho}) can be put into a matrix form (in the
stochastic $2-$dimensional space),
\begin{align}
\partial_t\widetilde{\rho}(t)=\bigl [-i\bigl
(\mathcal{H}^\times_{0}\widetilde{1}-{\cal
V}^\times\widetilde{F}\bigr )+\widetilde{W} \bigr
]\widetilde{\rho}(t)\ ,\label{eom}
\end{align}
where $\widetilde{1}$ is the $2\times 2$ unit matrix, while
\begin{align}
\widetilde{W}=\left(\begin{array}{cc}
 -w^{}_{-+} & w^{}_{+-} \\
 w^{}_{-+} & -w^{}_{+-}
\end{array}\right)\equiv \lambda(\widetilde{\cal T}-\widetilde 1)
\end{align}
is the relaxation matrix [see Eq. (\ref{wp})].   In the above we
have used the notations
\begin{align}
\widetilde{F}=\left(\begin{array}{cc}
 1 & 0 \\
 0 & -1
\end{array}\right)\ ,\ \ \ \widetilde{\cal T}\equiv \left(\begin{array}{cc}
 p^{}_+ & p^{}_+ \\
 p^{}_- & p^{}_-
\end{array}\right)\ .
\end{align}
Note that Eq. (\ref{eom}) contains two types of operators: ${\cal
H}^\times_0$ and $ {\cal V}^\times$ are Liouville super-operators
(represented by $9\times 9$ matrices), while ${\widetilde F}$ and
${\widetilde W}$ are $2\times 2$ matrices acting in the
$2-$dimensional space of the stochastic variable. The matrix
${\widetilde \rho}$ is of order $18\times 18$.

Using a Laplace transform,
$\widetilde{\rho}(s)=\int_{0}^{\infty}dt
e^{-st}\widetilde{\rho}(t)$ (unless specifically stated, all
$\widetilde\rho$'s below depend on $s$), Eq. (\ref{eom}) becomes
\begin{align}
&(s\widetilde{1}-\widetilde{W})(\widetilde\rho^{}_{11}\pm\widetilde\rho^{}_{22})=R^\pm\widetilde{1}+2J{\rm Im}[\widetilde\rho^{}_{13}\pm\widetilde\rho^{}_{23}]\ ,\nonumber\\
&(s\widetilde{1}-\widetilde{W})\widetilde\rho^{}_{12}=\rho^{}_{12}(0)-iJ(\widetilde\rho^{}_{13}-\widetilde\rho^{}_{32})\ ,\nonumber\\
&[s\widetilde{1}-\widetilde{W}-i\widetilde e]\widetilde\rho^{}_{13}=iJ(\widetilde\rho^{}_{33}-\widetilde\rho^{}_{11}-\widetilde\rho^{}_{12})\ ,\nonumber\\
&[s\widetilde{1}-\widetilde{W}-i\widetilde
e]\widetilde\rho^{}_{23}=iJ(\widetilde\rho^{}_{33}-\widetilde\rho^{}_{22}-\widetilde\rho^{}_{21})\
,\label{eqrho}
\end{align}
where we have set $\rho^{}_{13}(0)=\rho^{}_{23}(0)=0$ and defined
$\widetilde e\equiv\epsilon\widetilde{1}-\zeta\widetilde F$. Using
the last two equations,
and defining
\begin{align}
&\widetilde
A=[s\widetilde{1}-\widetilde{W}-i\widetilde e]^{-1}\equiv \widetilde A^{}_R+i\widetilde A^{}_I\ ,
\end{align}
we find
${\rm
Im}[\widetilde\rho^{}_{13}+\widetilde\rho^{}_{23}]=J\widetilde
A^{}_R \widetilde B$, where 
$\widetilde B \equiv
2\widetilde\rho^{}_{33}-\widetilde\rho^{}_{11}-\widetilde\rho^{}_{22}-2{\rm
Re}[\widetilde\rho^{}_{12}]$. 
Substitution into Eq. (\ref{eqrho}) then yields
\begin{align}
&(s\widetilde 1-\widetilde
W)(\widetilde\rho^{}_{11}+\widetilde\rho^{}_{22})=R^+\widetilde{1}+2J^2\widetilde
A^{}_R \widetilde B\ ,\nonumber\\
&(s\widetilde 1-\widetilde W){\rm
Re}[\widetilde\rho^{}_{12}]=y\widetilde{1}+J^2\widetilde
A^{}_R\widetilde B\ .\label{rho1122}
\end{align}

It is easy to check that $\sum_n\rho^{}_{nn}(t)\equiv 1$ implies
that
$(s\widetilde{1}-\widetilde{W})\sum_n\widetilde\rho^{}_{nn}=1$,
i.e. $\sum_n\widetilde\rho^{}_{nn}=\widetilde P$, where
$\widetilde P\equiv(s\widetilde 1-\widetilde W)^{-1}$ is the
Laplace transform of the probabilities to start at $a$ and end up
at $b$ at time $t$. Combining this identity with the above equations, one has
\begin{align}
\widetilde B=\bigl (2-3R^+-2y\bigr )[s\widetilde 1-\widetilde
W+8J^2\widetilde A^{}_R]^{-1}.
\end{align}
To complete  the solution of the Liouville equations we also find
two coupled equations for
$\widetilde\rho^{}_{11}-\widetilde\rho^{}_{22}$ and ${\rm
Im}[\widetilde\rho^{}_{12}]$:
\begin{align}
&\widetilde\rho^{}_{11}-\widetilde\rho^{}_{22}=\widetilde C\bigl (R^-\widetilde{1}+4J^2\widetilde A_I\widetilde Z{\rm Im}[\rho^{}_{12}(0)]\bigr )\ ,\nonumber\\
&{\rm Im}[\widetilde\rho^{}_{12}]=\widetilde Z\bigl
({\rm Im}[\rho^{}_{12}(0)]-J^2\widetilde A_I(\widetilde\rho^{}_{11}-\widetilde\rho^{}_{22})\bigr )\ ,\nonumber\\
&\widetilde Z=(s\widetilde 1-\widetilde W+2 J^2 \widetilde
A_R)^{-1}\ ,
\nonumber\\
&\widetilde C=(\widetilde Z^{-1}+4 J^4\widetilde A_I\widetilde
Z\widetilde A_I)^{-1}\ .
\end{align}

The time evolution of the density matrix follows inverse Laplace
transforms, which involve decaying oscillations, approaching
asymptotic time independent limits as $t\rightarrow\infty$.
Before presenting an example of this time evolution, we consider these latter limits,
using the identity
$\widetilde{\rho}(t\rightarrow\infty)=\lim_{s\rightarrow
0}[s\widetilde\rho(s)]$. Noting that $\widetilde A_I,~\widetilde
Z$ and $\widetilde C$ all remain finite as $s\rightarrow 0$, we
conclude that
$\widetilde\rho^{}_{11}(\infty)-\widetilde\rho^{}_{22}(\infty)=\Im[\widetilde\rho^{}_{12}(\infty)]=0,~  
\widetilde\rho^{}_{13}(\infty)=\widetilde\rho^{}_{23}(\infty)=0$.
Tedious but straightforward algebra is
needed to show that
\begin{align}
&\lim_{s\rightarrow 0}J^2 \widetilde A_R[s\widetilde 1-\widetilde W+8J^2\widetilde A^{}_R]^{-1}\nonumber\\
&\ \ \ \ = \frac{1}{16}\left (\begin{array}{cc}1+\epsilon/\zeta & 1+\epsilon/\zeta \\
 1-\epsilon/\zeta & 1-\epsilon/\zeta\end{array}\right )\ .
\end{align}
Using also  $\widetilde P(t\rightarrow \infty)=\lim_{s\rightarrow
0}[s\widetilde P]= \widetilde{\cal T}$, we find that the matrix
$\widetilde \rho$  splits as
\begin{align}
(b|\rho^{}_{mn}|a)=\rho_{mn}^{\infty}(b|\widetilde{\cal
T}|a)=\rho_{mn}^{\infty}p^{}_b\ ,\label{split}
\end{align}
with
\begin{align}
\rho^{\infty}=\frac{1}{8}\left [\begin{array}{ccc}R^++2-2y & 6y+2-3R^+ &0 \\
 6y+2-3R^+ & R^++2-2y & 0\\
 0 & 0 & 4-2R^++4y\end{array}\right ].\label{rhoinf}
 \end{align}
Since the factor $p^{}_b$ in (\ref{split}) gives the probability
to find the stochastic variable in state $b$, it does not affect
the actually measured density matrix, which will always approach
$\rho^{\infty}$. In particular, $[\widetilde{\cal T}]^{}_{av,a}=1$
and therefore all the averages will yield the same results. When
the electron is initially placed on the qubit, i.e.  $R^+=1$, then
$\rho^\infty $ depends only on $y$, and {\it not} on the other
details of the initial state. The result is also independent of
the non-stochastic Hamiltonian ($\epsilon,~J$) and of the
stochastic noise ($\Delta p,~\zeta,~\lambda$), as long as
$\lambda>0$ and $|\Delta p|<1$. In contrast, when $\Delta p=\pm
1$, i.e. at zero temperature, the system jumps to the state with
$\epsilon^{}_3=\epsilon\mp\zeta$ and then it stays there forever,
behaving like the non-stochastic system with that energy.

Substituting Eq. (\ref{rhoinf}) with $R^+=1$ into the decoherence
measures (\ref{DD}) and (\ref{SS}) yields maximal decoherence,
$\rho^{}_{nm}(\infty)=\delta^{}_{nm}/3$, when $y=1/6$.
When  the electron starts in state $|1\rangle$, then $y=0$, and
one has asymptotic partial decoherence,
$\rho^{}_{11}=\rho^{}_{22}=3/8,~\rho^{}_{12}=-1/8$.
When the initial state of the qubit is fully entangled,
$(|1\rangle+e^{i\alpha}|2\rangle)/\sqrt{2}$, then
\begin{align}
\rho(0)=\frac{1}{2}\left [\begin{array}{ccc}1 & e^{-i\alpha} &0 \\
 e^{i\alpha} & 1 & 0\\
 0 & 0 & 0\end{array}\right ]\ ,
 \end{align}
 so that $y=\frac{1}{2}\cos\alpha$ and
 $\rho_{33}^\infty=\frac{1}{2}\cos^2(\alpha/2)$. Measuring this occupation of the control register then
 yields a measure of the phase $\alpha$,
 in spite of the decoherence!

Both measures of decoherence, Eqs. (\ref{DD}) and (\ref{SS}),
vanish at all times if the electron starts with $y=-1/2~
(\alpha=\pi)$, which corresponds to the ``anti-bonding" state
$|AB\rangle=(|1\rangle-|2\rangle)/\sqrt{2}$. Being fully
antisymmetric under $1\leftrightarrow 2$, $|AB\rangle$ does not
couple to $|3\rangle$, and is not affected by the symmetric
Hamiltonian. Therefore, this state forms the {\it decoherent free
subspace}, which remains constant and fully protected at all
times. As $y$ increases from $-1/2$, one encounters a mixture of
this ``anti-bonding" state and the other two orthogonal
eigenstates of ${\cal H}_0$, and thus the system develops  various
degrees of partial decoherence.   The value $y=1/2~(\alpha=0)$
corresponds to the initial ``bonding" state,
$|B\rangle=(|1\rangle+|2\rangle)/\sqrt{2}$. In this case,
$\rho^{\infty}_{11}=\rho^{\infty}_{22}=\rho^{\infty}_{12}=1/4$,
and the projection of the quantum state onto the qubit subspace
remains in the fully coherent state $|B\rangle$.

\begin{widetext}

\begin{figure}[ht]
\includegraphics[width=5.5cm]{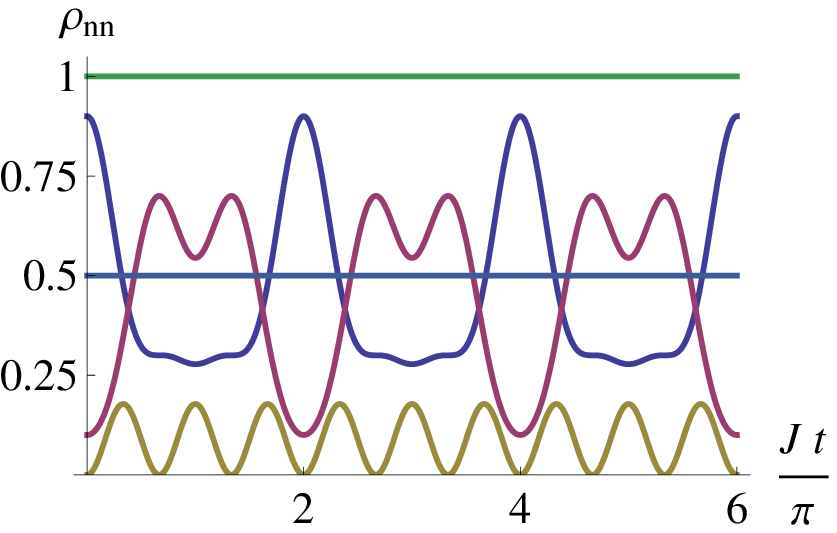}\ \ \
\includegraphics[width=5.5cm]{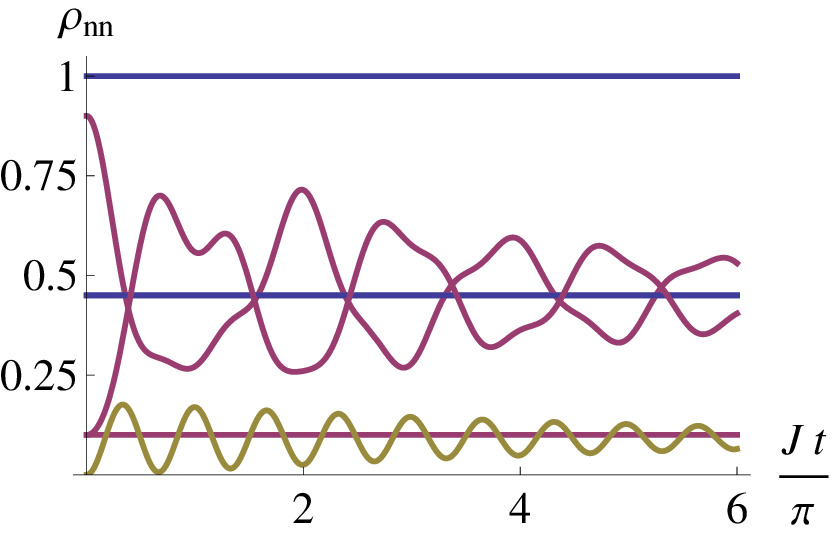}\ \ \
\includegraphics[width=5.5cm]{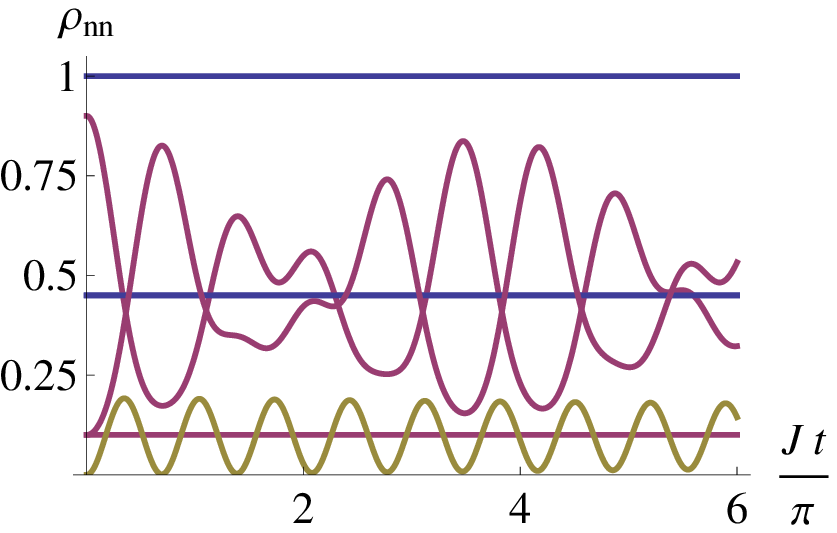}
\caption{The site occupations for an initial coherent state with
$y=-0.3$, and $\epsilon=J$. Left: non-stochastic limit. Center and
right: average occupations for stochastic cases, with
$\zeta=\lambda=0.5J$, $\Delta p=.1$ and $.9$. $\rho^{}_{11}$ and
$\rho^{}_{22}$ approach $(3-2y)/8=0.45$ and $\rho^{}_{33}$
approaches $(1+2y)/4=0.1$ (shown by horizontal lines).} \label{1}
\end{figure}

\end{widetext}

\begin{figure}[ht]
\begin{center}
\includegraphics[width=4 cm]{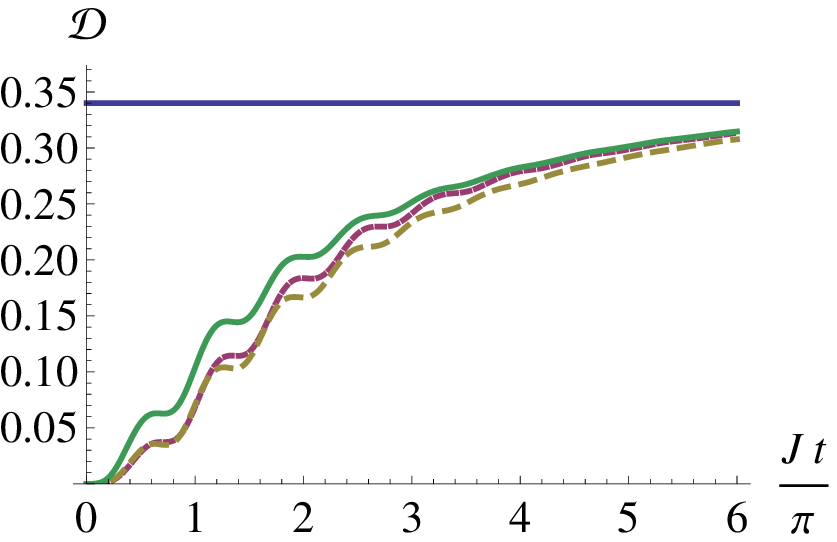}\ \
\includegraphics[width=4 cm]{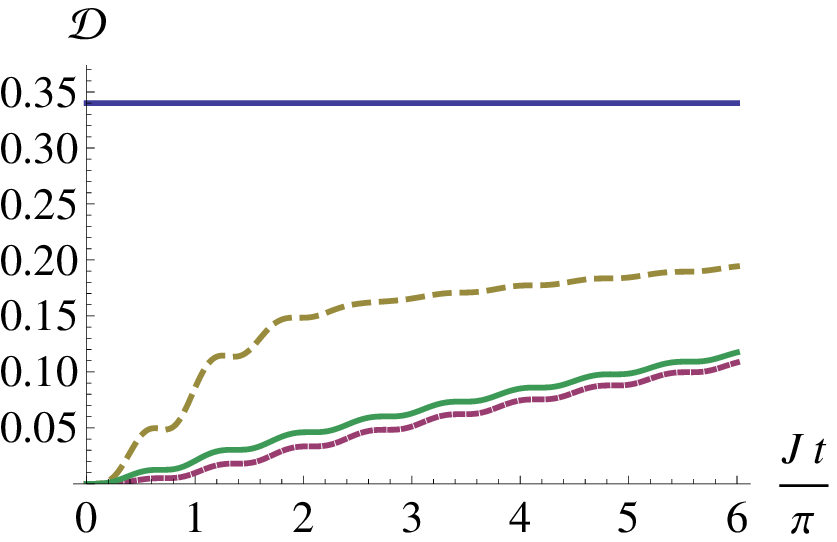}
\end{center}
 \caption{Time dependence of the decoherence measure  ${\cal D}$, for the same parameters as in Fig.
 \ref{1}. Left (right): $\Delta p=0.1~(0.9)$
 and with $\zeta=\lambda=0.5J$. Small (large) dashes correspond to $[\rho]^{}_{av,+}$ ($[\rho]^{}_{av,-}$). Full lines
 correspond to $[\rho]^{}_{av}$.
 The horizontal line shows the asymptotic value, ${\cal D}=0.34$.}\label{2}
\end{figure}

To demonstrate typical results, we start with an arbitrary choice
of parameters, $y=-0.3,~R^+=1,~R^-=\sqrt{1-4y^2}$ and set
$\epsilon=J$. Figure \ref{1} shows the three occupations
$\rho^{}_{nn}(t)$. The non-stochastic limit  
(left panel) exhibits Rabi oscillations with period $2\pi/J$,
corresponding to the difference between eigen-energies of ${\cal
H}^{}_0$; the coupling of the qubit to the control register causes
periodic partial periodic occupation of the latter.  In the
stochastic cases, small values of $\lambda(\ll J)$ and/or
$\zeta(\ll J)$ yield small corrections to this non-stochastic
picture at short times, with an eventual ``flow" to the asymptotic
behavior described above. The remaining panels of Fig. \ref{1}
show $[\rho]^{}_{av}$ for a relatively large stochasticity,
$\zeta=\lambda=J/2$ and $\Delta p=p^{}_+-p^{}_-=0.1$ (high $T$) or
$0.9$ (low $T$). The other averages $[\rho]^{}_{av,\pm}$ exhibit
similar qualitative behavior, and all the averages approach the
asymptotic values given by Eq. (\ref{rhoinf}). Clearly, the
decoherence time is longer for the lower temperature (larger
$|\Delta p|$). Figure \ref{2} shows the decoherence measure ${\cal
D}$, calculated using the three averages $[\rho]^{}_{av,\pm}$
(dashed lines) and $[\rho]^{}_{av}$ (full lines).  Note also that
the asymptotic limit ${\cal D}=0.34$ is much smaller than the
fully decoherent value $2/3$, demonstrating the partial
decoherence. Also, the approach to this asymptotic value is much
slower for the lower temperature.

The equations of motion become more complicated when one adds
direct hopping between the two qubit states, i.e. $J^{}_{12}\ne
0$. We have solved the equations of motion for this generalized
case, by inverting the full $9\times 9$ matrices for the
super-operators \cite{tobe}. Surprisingly, the asymptotic result
(\ref{rhoinf}) remains the same, independent of $J^{}_{12}$. This
robust result depends only on the symmetry $1 \leftrightarrow 2$.
We have also checked that any deviation from this symmetry (e.g.
by introducing a Aharonov-Bohm flux inside the ring formed by the
three sites, or by having different energies on sites 1 and 2)
immediately changes the asymptotic limit back into the fully
decoherent one.

It would be very interesting to check if Eq. (\ref{rhoinf}) also
holds for other sources of noise, e.g. a capacitive coupling to a
point contact \cite{gur1}. It would also be interesting to
consider generalizations of these results to systems containing
several qubits. We hope that our results will also be tested
experimentally.

SD is grateful to BGU for the warm hospitality he received during
several visits, when this project was carried out. AA and OEW
acknowledge discussions with S. Gurvitz and with Y. Imry and support from the DIP.

\end{document}